# The role of nonlinear optical absorption in narrow-band difference-frequency terahertz-wave generation

Matteo Cherchi,[1,2,a)] Saverio Bivona,[1,2] Alfonso C. Cino,[3] Alessandro C. Busacca,[1,3] and Roberto L. Oliveri[1,2]

[1]*CNISM – Consorzio Nazionale Interuniversitario per le Scienze Fisiche della Materia, Unità di ricerca di Palermo, Università di Palermo, Piazza Marina 57, I-90133 Palermo*

[2]*DIFTER – Dipartimento di Fisica e Tecnologie Relative, Università di Palermo, viale delle Scienze edificio 18, I-90128 Palermo, Italy*

[3]*DIEET- Dipartimento di Ingegneria Elettrica, Elettronica e delle Telecomunicazioni, Università di Palermo, viale delle Scienze edificio 9, I-90128 Palermo, Italy*

We present a general analysis of the influence of nonlinear optical absorption on terahertz generation via optical difference frequency generation, when reaching for the quantum conversion efficiency limit. By casting the equations governing the process in a suitably normalized form, including either two-photon- or three-photon-absorption terms, we have been able to plot universal charts for phase matched optical-to-terahertz conversion for different values of the nonlinear absorption coefficients. We apply our analysis to some experiments reported to date, in order to understand to what extent multiphoton absorption could have played a role and also to predict the maximum achievable conversion efficiency at higher peak pump intensities.

a) cherchi@unipa.it

# Introduction

Off-laboratory applications of terahertz radiation demand efficient table-top sources, which can rely either on ultra high frequency microwave circuits or optical-to-terahertz converters [1]. In particular Difference Frequency Generation (DFG) is one of the most promising physical mechanism to generate terahertz radiation from optical sources [2]. It exploits the quadratic nonlinear susceptibility of quadratic nonlinear materials [3] to convert optical pump photons with frequency $\omega_u$ into optical signal photons with frequency $\omega_v < \omega_u$ and terahertz photons of frequency $\omega_w = \omega_u - \omega_v$. High conversion efficiencies typically require optical intensities at least of the order of tens of $GW/m^2$. This implies that, in general, also other nonlinear effects can be involved in these experiments. In particular in this work we propose a universal approach to analyze the detrimental effects of two- and three-photon absorption on terahertz photon conversion efficiency, in order to provide a reference for past and future experiments with high intensity peak lasers.

Among the different optical-to-terahertz conversion mechanisms [4] (including optical rectification, photoconduction, Cherenkov radiation), the only one that is scalable both with pump power and sample length, is phase matched DFG based on narrow band optical pulses and pump photon energies below the bandgap of the chosen nonlinear material. This means that DFG is probably the only method available to approach the quantum efficiency limit. As a matter of fact, the only experimental results approaching the quantum limit reported to date [5] are based on this approach.

Previous analyses and experiments highlighting the nonlinear absorption (NLA) effects in optical-to-terahertz generation [6-8] have been focused on optical rectification of ultrafast laser pulses converted into broadband terahertz pulses, that is DFG within each pulse. In this kind of non-scalable experiments, the pulse length limits the sample length. This is why, even using very high pump intensities, the quantum limit has never been approached, also because of the detrimental effects of NLA on very high peak intensity broadband optical pulses. Instead, we present, for the first time in the literature, a detailed analysis of the role of NLA on DFG, in the standard narrow band pump-signal configuration. In particular we aim to estimate its detrimental effects when very high peak optical intensities are launched in the attempt to attain the quantum conversion limit.

## Physical model

In two recent papers [9], [10] we have addressed the problem of the analysis of the quantum efficiency limit, a regime that requires the exact numerical solution of the coupled equations [11] governing the conversion process in the frame of couple mode theory. In particular in Ref. [10] we proposed the use of the universal charts defined as follows. All DFG experiments can be effectively described by a scalar model, by defining a suitable effective nonlinear coefficient $d_{eff}$ [3] and an effective area $A_{DFG} \equiv \langle e_w | e_w \rangle \langle e_v | e_v \rangle \langle e_u | e_u \rangle / \langle e_w e_v | e_u \rangle^2$, that is the inverse of the overlap integral of the spatial distributions $e_q(x,y)$ of the three waves [12], which are assumed to not vary along propagation. In this way it is possible to define a Figure of Merit [9] (FOM) $\mathcal{F} \equiv \xi^2/\alpha^2$ (having the dimensions of the inverse of a photon flux) featuring the terahertz absorption coefficient $\alpha$ and the coupling coefficient $\xi \equiv d_{eff}[2Z_0 \hbar \omega_w \omega_u \omega_v / (c^2 n_w n_u n_v A_{DFG})]^{1/2}$, where $c$ is the vacuum speed of light, $\hbar$ is the Planck constant, $Z_0$ is the vacuum impedance, $n_q$ ($q = u, v, w$) are the refractive indexes for the three waves This enables to write the coupled equations in terms of the normalized distance $\zeta \equiv z\alpha$, of the normalized momentum mismatch $\kappa \equiv 2\Delta k/\alpha$ and of the normalized photon fluxes amplitudes $\hat{q}(z;t) \equiv \sqrt{\mathcal{F}} q(z;t)$, as follows

$$\begin{cases} \dfrac{d\hat{w}}{d\zeta} = -i\hat{u}\hat{v}^* - \dfrac{1-i\kappa}{2} \hat{w} \\ \dfrac{d\hat{v}}{d\zeta} = -i\hat{u}\hat{w}^* \\ \dfrac{d\hat{u}}{d\zeta} = -i\hat{v}\hat{w} \end{cases} \qquad (1)$$

Here we are assuming terahertz absorption lengths much longer than terahertz wavelengths [13], negligible optical losses, and pulse durations not too smaller than the time of flight in the system, in order to avoid the effects of group velocity dispersion. The space-time dependent $q(z;t)$ functions are assumed to be slowly varying functions of $z$ and are normalized such that their square moduli are the photon fluxes $N_q$ (number of photons per unit time) of each wave. In the case of free space experiments, equations (1) can reliably model nonlinear conversion only if the terahertz beam profile can be assumed to be almost constant along the whole crystal length $L$, that is if the waists of the optical beams are not too much smaller than $r_R \equiv \sqrt{cL/(n_w \omega_w)}$. Otherwise, these equations overestimate terahertz conversion, so that their predictions can provide just an upper limit to maximum conversion efficiency [14,15]. Even though this could seem a strong limitation of our formalism, we notice that tight focusing can be a convenient way only to slightly improve very small conversion efficiencies, but cannot be an effective way to approach the quantum conversion limit. This is because terahertz beam diffraction would strongly affect the typical $L^2$ scaling up of photon conversion. A terahertz Rayleygh length $l_R \ll L$ would act as a cut-off length leading to the scaling law $L \times l_R$ [15], and the quantum limit could be hardly approached. As a matter of fact, the only experimental results approaching the quantum limit reported to date [5] has been obtained with loosely focused optical beams.

We point out that all assumptions leading to equations (1) are meaningful, since they all correspond to the optimal conditions for narrow-band DFG, and their fulfilment is mandatory when reaching for the quantum efficiency limit. As a validation of the proposed model, we also point out that, based on equations (1), we have been able to provide a physically reliable interpretation [10] of the high conversion efficiency results of Ref. [5], which were obtained close to complete pump depletion. This led us also to a reliable estimation of the nonlinear coefficient of high quality GaSe crystals.

Thanks to the dimensionless form of equations (1), the number of independent variables for terahertz generation (i.e. with initial terahertz photon flux $N_{w0} = 0$) is reduced to four. They are: the initial normalized pump photon flux $\hat{N}_{u0} \equiv |u_0|^2 = \mathcal{F} N_{u0}$, the ratio $R \equiv N_{v0}/N_{u0}$ between the initial signal and pump photon fluxes, the normalized phase mismatch $\kappa$, and the normalized propagation distance $\zeta$. By fixing a constraint to any two of these variables, it is possible to plot universal charts for the terahertz photon conversion efficiency $\eta(N_{u0},R,\kappa,\zeta) \equiv N_w/N_{u0}$ as a family of curves that are all functions of one of the two unconstrained variable, each curve corresponding to a different value of the other unconstrained variable, acting as a free parameter. For the sake of practice, it is also convenient to introduce the reference powers $\bar{P}_q \equiv \hbar\omega_q/\mathcal{F}$, and the reference intensities $\bar{I}_q \equiv \bar{P}_q/A_{\text{DFG}}$ in order to define the normalized powers and intensities $\hat{P}_{q0} \equiv P_{q0}/\bar{P}_q = \hat{I}_{q0} \equiv I_{q0}/\bar{I}_q = \hat{N}_{q0}$ (notice that, by definition, $\bar{P}_q$ are proportional to $A_{\text{DFG}}$, whereas $\bar{I}_q$ are independent of $A_{\text{DFG}}$). Physically, $\bar{P}_q$ and $\bar{I}_q$ estimate the optical fluence regime needed to approach the quantum conversion limit.

## Multiphoton absorption terms

The universal charts presented in our previous work [10] showed how the proposed normalization allows to keep the number of degrees of freedom to a minimum and to analyze the contributions of every meaningful physical parameter in a very general way. In this paper we will show how these charts can be extended to include other nonlinear effects that can compete with DFG, especially when launching high optical intensities. In particular we will analyze the detrimental effects of two-photon absorption (2PA) and three-photon absorption (3PA) [16], that are due to the imaginary parts of the third- and fifth-order susceptibility respectively. In fact, even though optical linear absorption can be avoided working in the near-infrared transparency region of most common nonlinear materials, in general 2PA and 3PA can be avoided only working with optical pump and signal in the low energy tail of the mid-infrared region, which is not covered by the most common laser sources. Since 2PA scales linearly with the light intensity and 3PA scales with the square of the light intensity, in general, they cannot be neglected when using high peak optical intensities.

From the transition diagrams of bichromatic 2PA (Fig. 1.g), it is straightforward to calculate the normalized terms [17] that must be added to the last two of equations (1) in order to take into account optical 2PA, that are

$$\begin{cases} 2\text{PA}_{\hat{v}} = -\frac{1}{2}\left(\hat{\beta}_{vv}|\hat{v}|^2 + 2\hat{\beta}_{vu}|\hat{u}|^2\right)\hat{v} \\ 2\text{PA}_{\hat{u}} = -\frac{1}{2}\left(\hat{\beta}_{uu}|\hat{u}|^2 + 2\hat{\beta}_{uv}|\hat{v}|^2\right)\hat{u} \end{cases}. \quad (2)$$

Here we have defined the dimensionless 2PA coefficients $\hat{\beta}_{pq} \equiv \overline{P}_q / A_{pq} \times \beta_p / \alpha$, where $\beta_p$ is the usual 2PA coefficient at frequency $\omega_p$ and $A_{pq} \equiv \langle e_p | e_p \rangle \langle e_q | e_q \rangle / \langle e_p e_q | e_p e_q \rangle$ are the respective 2PA effective areas. Since in typical experiments $\omega_u \approx \omega_v$ and all the effective areas are almost identical, in this paper we will always assume $\hat{\beta}_{pq} \equiv \hat{\beta}$ for all $p$ and $q$ combinations.

In Fig. 1.a and Fig. 1.b we show the effect of 2PA on the $R = 1$ curves for phase matched DFG, corresponding to the maximum condition [10], treating $\hat{\beta}$ as a parameter: $\eta_{\max}$ is the maximum achievable photon conversion efficiency at a given $\hat{P}_{u0}$, corresponding to a propagation distance $\zeta_{\max}$. From Fig. 1.a it is clear that the effect of increasing $\hat{\beta}$ at lower $\hat{P}_{u0}$ is to slightly diminishing $\zeta_{\max}$, while at higher $\hat{P}_{u0}$ it significantly increases $\zeta_{\max}$. For any $\hat{\beta} \neq 0$, $\eta_{\max}$ reaches a maximum value $\max(\eta_{\max})$ at a certain $\hat{P}_{u0,\max}$, and then start decreasing, as clearly shown in Fig. 1.b. This corresponds to a sub-linear growth of $\hat{P}_w$ with $\hat{P}_{u0}$. A similar behaviour is found for $R = 10^{-2}$, with the difference that, at lower $\hat{P}_{u0}$, $\zeta_{\max}$ is almost unchanged, as shown in Fig. 1.c. showing also that lower slopes correspond to higher $\hat{P}_{u0} \times \hat{\beta} \times \zeta_{\max}$ values. Furthermore, it is worth noting that, in this case, when $\hat{\beta} \geq 10^{-1}$ conversion efficiency is significantly affected also at lower $\hat{P}_{u0}$, as clearly shown in Fig. 1.d. This is basically due to the amplification dynamics of the optical signal. In Fig. 1.e it is shown that, for any given $R$ value, $\hat{P}_{u0,\max}$ roughly scale as $1/\hat{\beta}$, in agreement with the linear dependence on $\hat{P}_u$ and $\hat{P}_v$ of the 2PA terms (2). From Fig. 1.f it is clear that the corresponding max($\eta_{max}$) is significantly affected only when $\hat{\beta} \geq 10^{-2}$.

We now focus on 3PA, whose additional terms can be easily calculated [17] from the relative transition diagrams (Fig. 2.g) to be

$$\begin{cases} 3PA_{\hat{v}} = -\frac{1}{2}\left(\hat{\gamma}_{vvv}|\hat{v}|^4 + 6\hat{\gamma}_{vvu}|\hat{v}|^2|\hat{u}|^2 + 3\hat{\gamma}_{vuu}|\hat{u}|^4\right)\hat{v} \\ 3PA_{\hat{u}} = -\frac{1}{2}\left(\hat{\gamma}_{uuu}|\hat{u}|^4 + 6\hat{\gamma}_{uuv}|\hat{u}|^2|\hat{v}|^2 + 3\hat{\gamma}_{uvv}|\hat{v}|^4\right)\hat{u} \end{cases}. \tag{3}$$

Here we have defined the dimensionless 3PA coefficients $\hat{\gamma}_{pqr} \equiv \bar{P}_q \bar{P}_r / A_{pqr}^2 \times \gamma_p / \alpha$, in terms of the standard 3PA coefficient $\gamma_p$, and of the 3PA effective areas $A_{pqr} \equiv \left(\langle e_p|e_p\rangle\langle e_q|e_q\rangle\langle e_r|e_r\rangle / \langle e_p e_q e_r|e_p e_q e_r\rangle\right)^{1/2}$. Since in most common experimental configurations $\omega_u \approx \omega_v$ and all the effective areas are almost identical, in this paper we will always assume $\hat{\gamma}_{pqr} = \hat{\gamma}$ for all $p$, $q$, and $r$ combinations. In this case $\zeta_{max}$ for phase matched DFG is almost unchanged at lower $\hat{P}_{u0}$, in both regimes $R = 1$ and $R = 10^{-2}$ (Fig. 2.a and Fig. 2.c). Also, when $\hat{P}_{u0}^2 \times \hat{\gamma} \times \zeta_{max} > 1$, in the former regime $\zeta_{max}$ becomes almost independent of $\hat{P}_{u0}$, whereas in the second regime it starts growing after a minimum value. Again, for any $\hat{\gamma} \neq 0$, $\eta_{max}$ reaches a maximum value max($\eta_{max}$) at a certain $\hat{P}_{u0,max}$ and then start decreasing, as clearly shown in Fig. 2.b and Fig. 2.d. In terms of $\hat{P}_w$, this efficiency decrease corresponds to a plateau when $R = 1$ and to a very slow decrease when $R = 10^{-2}$. This feature clearly distinguishes 2PA from 3PA. In Fig. 2.e it is shown that, for all $R$ values, $\hat{P}_{u0,max}$ roughly scale as $\hat{\gamma}^{-1/2}$, in agreement with the dependence on the products $\hat{P}_p \hat{P}_q$ of the 3PA terms (3). From Fig. 2.f it is clear that the corresponding max($\eta_{max}$) is significantly affected only when $\hat{\gamma} > 10^{-3}$.

It should also be noticed that optical NLA can generate free carriers which, in turn, can absorb terahertz radiation through free carrier absorption (FCA). As a first approximation this effect can be regarded as a constant addition term to linear absorption, as far as optical intensity is not dramatically affected by NLA itself. In this way it is still possible to use the proposed normalized charts with the replacement $\alpha \to \alpha + \alpha_{FCA}$, overestimating, in general, FCA effects. Otherwise the following additional FCA terms should be added to the first of equations (1)

$$\begin{aligned}
FCA2_w &\equiv -\tau_{eff}\sigma_w \left[ \frac{\bar{P}_v}{A_{wv}} \left( \frac{\hat{\beta}_{vv}|\hat{v}|^2}{2\hbar\omega_v} + \frac{2\hat{\beta}_{vu}|\hat{u}|^2}{\hbar(\omega_v+\omega_u)} \right)|\hat{v}|^2 + \right. \\
&\left. + \frac{\bar{P}_u}{A_{wu}} \left( \frac{\hat{\beta}_{uu}|\hat{u}|^2}{2\hbar\omega_u} + \frac{2\hat{\beta}_{uv}|\hat{v}|^2}{\hbar(\omega_u+\omega_v)} \right)|\hat{u}|^2 \right] \frac{\hat{w}}{2} \\
FCA3_w &\equiv -\tau_{eff}\sigma_w \left[ \frac{\bar{P}_v}{A_{wv}} \left( \frac{\hat{\gamma}_{vvv}|\hat{v}|^4}{3\hbar\omega_v} + \frac{6\hat{\gamma}_{vvu}|\hat{v}|^2|\hat{u}|^2}{\hbar(2\omega_v+\omega_u)} + \frac{3\hat{\gamma}_{vuu}|\hat{u}|^4}{\hbar(\omega_v+2\omega_u)} \right)|\hat{v}|^2 + \right. \\
&\left. + \frac{\bar{P}_u}{A_{wu}} \left( \frac{\hat{\gamma}_{uuu}|\hat{u}|^4}{3\hbar\omega_u} + \frac{6\hat{\gamma}_{uuv}|\hat{u}|^2|\hat{v}|^2}{\hbar(2\omega_u+\omega_v)} + \frac{3\hat{\gamma}_{uuv}|\hat{v}|^4}{\hbar(\omega_u+2\omega_v)} \right)|\hat{u}|^2 \right] \frac{\hat{w}}{2}
\end{aligned} \quad (4)$$

where $\sigma_w$ is the FCA cross-section at frequency $\omega_w$ and the effective time constant $\tau_{eff}$ depends on optical pulse durations and repetition rates, compared to the scattering time constant for the carriers through the interaction with phonons [6,18]. Similar terms could also account for free carrier absorption in the optical domain. Anyway we notice that, by suitably applying a voltage to the sample, it should be possible to reduce free carrier density up to two orders of magnitude [19].

As regards the possible role of nonlinear refraction in narrow-band DFG experiments, we notice that the phase matching acceptance bandwidth of centimetres long samples requires optical pulse durations longer than 100 ps. Since the typical fluence damage threshold values for nonlinear crystals are in the order of tens of kJ/m$^2$, this means that maximum launchable optical intensities are in the order of hundreds of TW/m$^2$. On the other hand, typical values of the nonlinear refractive indexes of the materials under consideration are in the order of 10$^{-5}$ m$^2$/TW [20], making optically induced refractive index changes always negligible.

## Discussion and applications

The quantum conversion limit in optical-to-terahertz generation has been approached for the first time [5] only recently. On the other hand, light intensities in all the other DFG experiments reported to date are far below the reference power $\bar{P}_u$ for pump depletion, so that the quantum conversion limit has not been approached. The aim of the proposed charts is to predict whether it will be possible, using higher optical peak intensities, to approach the quantum limit, even in presence of NLA.

As a practical example, we now apply our charts to the analysis of terahertz generation in GaAs and GaSe, two promising materials for terahertz generation, whose multiphoton absorption has been fully characterized in some recent papers [20], [21], [22]. In the following it should be taken into account that, unlike the experiments under consideration, the proposed charts correspond to the optimum length condition $\zeta = \zeta_{\max}(\hat{P}_{u0})$, and so they can just provide upper efficiency limits. Recently it has been reported relatively efficient generation ($\eta \approx 2\times 10^{-4}$, that is well below the quantum limit) of 2.2 THz radiation (corresponding to $\alpha \approx 3.66 \text{ cm}^{-1}$) in 5 mm long ($\zeta = 1.83 << \zeta_{\max}$) quasi phase matched GaAs samples [23], [24] with kWs peak power picoseconds pulses at wavelengths around 2.2 µm . The 80 µm waist optical beams were tightly focused with respect to the terahertz diffraction condition $r_R \approx 181$ µm, corresponding to a focusing parameter $\xi \equiv r_R^2/r_w^2 = 5.1$, so that, following the analysis of Ref. [15], our charts are expected to overestimate conversion efficiency by a factor of about 2.5. At the wavelengths under consideration, 2PA is well below threshold, and hence only 3PA should be kept into account. For this system $R \approx 1$, $\bar{P}_u \approx \bar{P}_v \approx 1.2$ MW and $\bar{I}_u \approx \bar{I}_v \approx 62$ TW/m$^2$, that means $\hat{P}_{u0} \approx 10^{-3}$, and, since $\gamma_u \approx \gamma_v \approx 0.3$ m$^3$/TW$^2$ and $\alpha \approx 3.7$ cm$^{-1}$, we find $\hat{\gamma} \approx 4$. From our charts, this clearly means that nonlinear absorption effects were completely negligible in those experiments. However, it could be wondered if, in similar experiments, it would be possible to approach the quantum limit by simply enhancing the launched peak intensity. From Fig. 2.e and Fig. 2.f, it is clear that, due to 3PA, increasing $\hat{P}_{u0}$, the maximum attainable efficiency in this system is around 10%, corresponding to about 100 kW optical peak power, that is still well below the optical damage threshold fluence of GaAs, which exceeds tens of kJ/m$^2$ [25,26] for nanosecond pulses. Actually, because of the tight focusing of the optical beam with respect to terahertz diffraction, the found result should be regarded as an upper limit to conversion efficiency. Using the same light intensity with much looser focusing can just help to reach this upper limit but cannot change the $\hat{\gamma}$ value, which, by definition, is invariant under scaling of the beam sizes (the same is true for $\hat{\beta}$). The only way to improve the quantum efficiency limit is to lower the terahertz absorption coefficient, that could be done, for example, by generating lower terahertz frequencies. In fact we point out that $\hat{\gamma} \propto \alpha^3 / \omega_w^2$, and so, moving for example to 1 THz (corresponding to $\alpha \approx 0.48$ cm$^{-1}$), we find $\hat{\gamma} \approx 4 \times 10^{-2}$. In this case, since $\bar{P}_u \propto \alpha^2 / \omega_w$, maximum achievable conversion efficiency in a $\zeta_{max} \approx 3$ cm long sample would be more than 40%, corresponding to $\hat{P}_{u0} \approx 0.8 \times \bar{P}_u \approx 36$ kW input peak power. Alternatively, the sample should be cooled to cryogenic temperatures, in order to reduce phonon absorption.

As regards GaSe, in a recent paper [5] $\eta = 39.2\%$ has been reported at 1.48 THz (corresponding to $\alpha \approx 0.2$ cm$^{-1}$), relying on birefringent phase matching in a 4.7 cm long ($\zeta = 0.94 < \zeta_{max} \approx 1.5$) crystal with hundreds of kW peak power optical nanoseconds pulses at wavelengths around 1 μm, where 2PA dominates ($\beta \approx 10$ m/TW). It should be noticed that, since in this experiment the waists of the optical pump and of the optical signal were quite different ($r_u = 0.75$ mm and $r_v = 1.93$ mm respectively), a rigorous analysis should rely on the exact calculation of the $\hat{\beta}_{pq}$ terms of Eq. (2). Anyway, in order to estimate the 2PA, we can approximate the Gaussian beam profiles with uniform distributions, to find $\bar{P}_u \approx \bar{P}_v \approx 50$ kW, $\bar{I}_u \approx \bar{I}_v \approx 28$ GW/m$^2$, $\hat{P}_{u0} \approx 4.6$, $R \approx 1/5$ and $\hat{\beta} \approx 10^{-2}$. The generated terahertz beam waist

$r_w = 1/\sqrt{1/r_u^2 + 1/r_v^2} = 0.70$ mm is comparable with the reference waist $r_R \approx 0.78$ mm (focusing parameter $\xi \equiv r_R^2/r_w^2 = 1.2$), and our charts are expected to overestimate the conversion efficiency by 20% only [15], that is within the experimental uncertainties. In the intensity regime of these experiments, 2PA is expected to reduce $\eta$ by 3% only, i.e. well within the experimental uncertainties. Hence in this case the effects of nonlinear absorption are less important than in GaAs and our charts predict that, in an optimal long sample ($\zeta_{max} \approx 0.4$), it should be possible to attain more than 70% photon conversion efficiency when launching 10 times higher optical intensities. This value is well below the typical value of damage threshold of high quality semiconductors, and can be certainly achieved with shorter (hundreds of picoseconds) pulses. We also notice that, since $\hat{\beta}$ scales as $\alpha/\omega_w$, the effects of 2PA should become appreciable, in the attempt to reach the quantum limit, when generating higher frequencies about 6 THz, corresponding to $\alpha \approx 10$ cm$^{-1}$.

## Conclusions

We presented a universal method to evaluate the role of NLA, when launching high peak optical intensities, in the attempt to achieve high terahertz conversion efficiencies in a phase matched DFG process. Our analysis show that the impact of NLA strongly depends on the amount of terahertz linear absorption and the higher the order of the multiphoton absorption process the stronger this dependence. In particular we have shown that, in general, nonlinear absorption in the optical domain cannot be neglected when the linear terahertz losses are too high. Remarkably, using some materials in some configurations, this can also make the quantum efficiency limit unachievable. On the other hand, unlike other optical-to-terahertz conversion mechanisms, in the best configurations DFG allows to approach the quantum efficiency limit even in presence of NLA.

M. Cherchi was supported by the Consorzio Nazionale Interuniversitario per le Scienze Fisiche della Materia (CNISM).

**Figure legends**

1. FIG. 1 Universal charts for phase matched DFG in presence of two photon absorption. a), c) optimum conversion length $\zeta_{max}$ vs. normalized pump peak power $\hat{P}_{u0}$ and b), d) corresponding maximum conversion efficiency for $R=1$ and $R=10^{-2}$. Each curve corresponds to a different value of the normalized 2PA coefficient $\hat{\beta}$. d), e) Optimum conversion power $\hat{P}_{u0,max}$ vs. $\hat{\beta}$ and corresponding maximum conversion efficiency max($\eta_{max}$). Each curve corresponds to a different value of the ratio $R$. The effects of FCA are neglected.

2. FIG. 2 Universal charts for phase matched DFG in presence of three photon absorption. Same as FIG. 1, but in terms of the 3PA coefficient $\hat{\gamma}$.

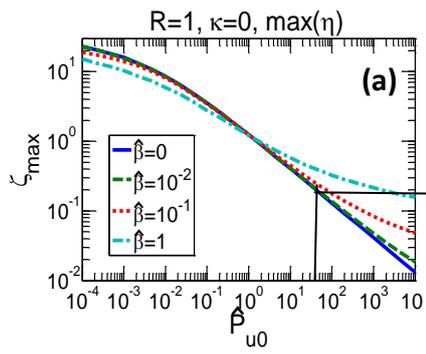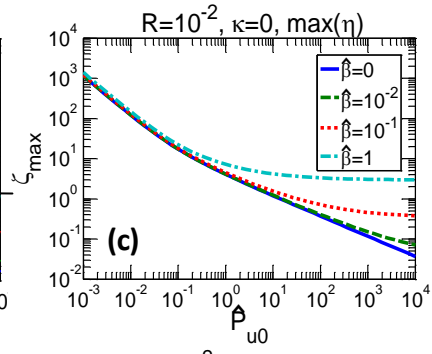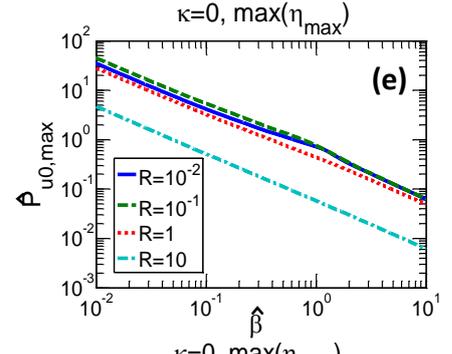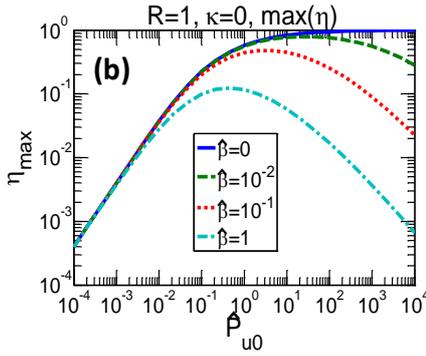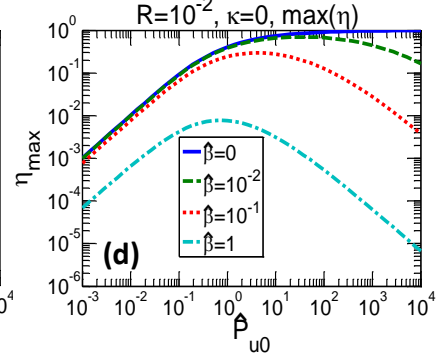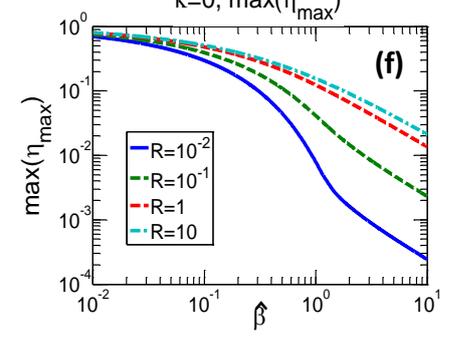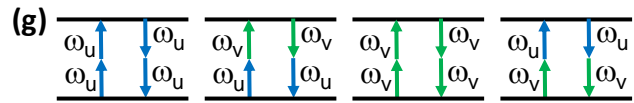

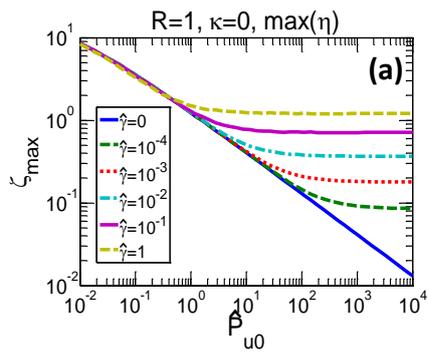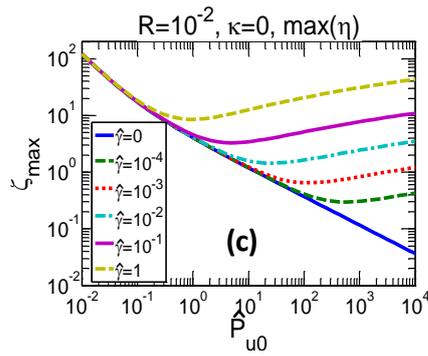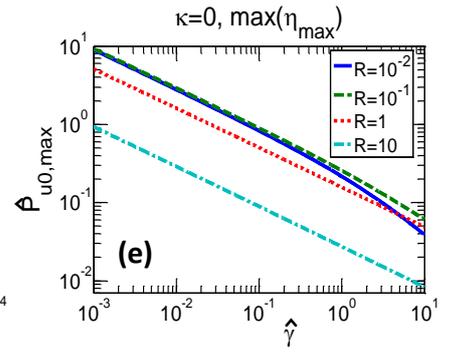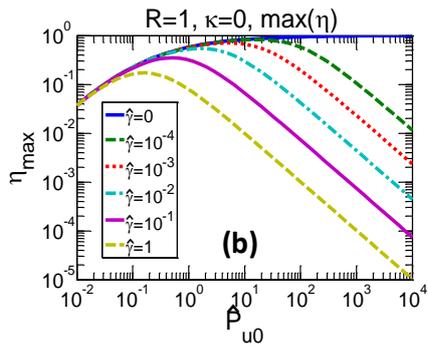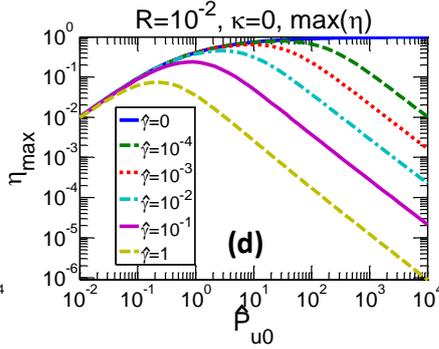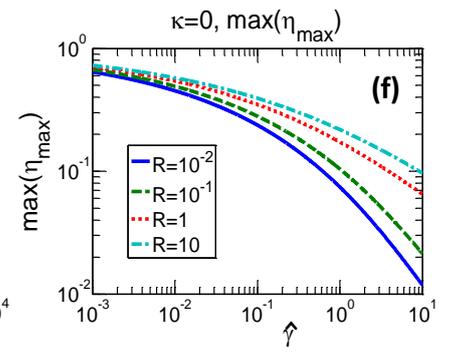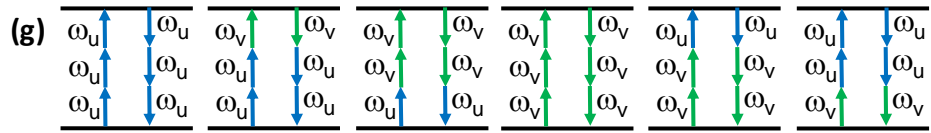